\documentstyle[11pt,epsfig]{article}
\newcommand{\etmiss}{E_T\!\!\!\!\!\!\!\!\! \not \,\,\,\,\,\,\,}
\newcommand{\ptmiss}{p_T\!\!\!\!\!\!\!\! \not \,\,\,\,\,\,\,}
\newcommand{\mtrec}{m_t^{\mathrm rec}}
\newcommand{\mwrec}{M_W^{\mathrm rec}}
\newcommand{\mzrec}{M_Z^{\mathrm rec}}

\begin{document}
\vspace*{-4cm}
\noindent
\hspace*{11cm}
UG--FT--98/99 \\
\hspace*{11cm}
hep-ph/9906462 \\
\hspace*{11cm}
June 1999 \\

\vspace{2cm}
\begin{center}
\begin{large}
{\bf $Zt$ and $\gamma t$ production via top flavour-changing \\
neutral couplings at the Fermilab Tevatron}
\end{large}

\vspace{0.5cm}
F. del Aguila, J. A. Aguilar--Saavedra \\
{\it Departamento de F\'{\i}sica Te\'{o}rica y del Cosmos \\
Universidad de Granada \\
E-18071 Granada, Spain} \\
~\\
Ll. Ametller \\
{\it Dep. F\'\i sica  i Enginyeria Nuclear \\
Universitat Polit\`ecnica de Catalunya \\
E-08034 Barcelona, Spain}\end{center}
\begin{abstract}
Associated single top production with a $Z$ boson or a photon at large hadron
colliders provides a precise determination of top flavour-changing neutral
couplings. The best way to measure these couplings with the up quark at Tevatron
is to search for events with three jets and missing energy or events with a
photon, a charged lepton, a jet and missing energy. Other decay
channels are also discussed.
\end{abstract}
\hspace{0.8cm} \noindent

\vspace{0.5cm}
Large hadron colliders will be top factories, allowing to measure its properties
very precisely. In contrast with its mass, which is the best known quark mass,
top couplings are very poorly known \cite{papiro1}. In this Letter we point out
that associated single top production with a $Z$ boson or a photon
is very sensitive to the flavour-changing neutral (FCN) couplings $Vtq$,
with $V$ a $Z$ boson, a photon or a gluon and $q$ a light quark $u$ or $c$. These
vertices are very small in the Standard Model (SM), being then an obvious place to
look for new physics.
Although top pair production gives large top samples, the leptonic
$Zt$ and $\gamma t$ signals become cleaner and statistically more
significant with increasing energy and luminosity. At any rate, the
determination of top FCN couplings from $Z t$ and $\gamma t$ production has a
comparable, if not a higher precision than from top decays as we will show here for
Tevatron. On the other hand, both
measurements have not only to be consistent but they will improve their
statistical significance when combined together. In the following
we concentrate on
Runs I and II with integrated luminosities of 109 pb$^{-1}$ and 2 fb$^{-1}$,
respectively. The charm contribution to associated top production with a $Z$
boson or a photon is 40 times smaller than the up contribution at Tevatron
energies $\sqrt s = 1.8-2$ TeV. Hence, Tevatron is only sensitive in these
production processes to top couplings with the up quark . This will not be the
case at the CERN Large Hadron Collider (LHC),
where the charm contribution becomes relevant. In any case there is no
model-independent reason for the $Vtu$ couplings to be small, and it is
theoretically important to measure them precisely.

The most significant decay channels depend on the collider and luminosity.
For $Zt$ production,
the $\nu \bar \nu jjb$ decay channel, with 
$Z \to \nu \bar \nu$ and $W \to q \bar q'$, gives the
best determination of the $Ztu$ couplings in both Tevatron runs. 
The $jj l \nu b$ mode, with $Z \to q \bar q$ and $W \to l \nu$, has a similar
branching ratio but a larger background.
For higher luminosity at Tevatron Run III
or at the LHC, the leptonic mode $l^+ l^- l' \nu b$, with both $Z$ and $W$
decaying leptonically, and the $b \bar b l \nu b$ mode are more significant.
For $\gamma t$ production the $\gamma l \nu b$ channel, with
$W \to l \nu$, gives a cleaner signal than the $\gamma j j b$ mode with
$W \to q \bar q'$, allowing a better determination of the $\gamma tu$ vertex.
This leptonic channel also gives the best limit on the $gtu$ coupling.
A detailed discussion of all these decay channels at LHC will be
presented elsewhere \cite{papiro10}.
Throughout this Letter we consider that the top quark decays predominantly into
$Wb$ \cite{papiro2}, and we sum $t$ and $\bar t$ production.

The Lagrangian involving FCN couplings
between the top, a light quark $q=u,c$ and a $Z$ boson, a photon $A$ or a gluon
$G^a$ can be written in standard notation as \cite{papiro4}
\begin{eqnarray}
{\mathcal L} & = & \frac{g_W}{2 c_W} \bar t \gamma_\mu (X_{tq}^L P_L +
 X_{tq}^R P_R) q Z^\mu \nonumber \\
& & + \frac{g_W}{2 c_W} \bar t (\kappa_{tq}^{(1)}-i \kappa_{tq}^{(2)} \gamma_5)
\frac{i \sigma_{\mu \nu} q^\nu}{m_t} q Z^\mu  \nonumber \\
& &  + e \bar t (\lambda_{tq}^{(1)}-i \lambda_{tq}^{(2)} \gamma_5)
\frac{i \sigma_{\mu \nu} q^\nu}{m_t} q A^\mu \nonumber \\
& & + g_s \bar t (\zeta_{tq}^{(1)}-i \zeta_{tq}^{(2)} \gamma_5)
\frac{i \sigma_{\mu \nu} q^\nu}{m_t}
T^a q G^{a\mu}+{\mathrm h.c.}\,, \label{ec:1}
\end{eqnarray}
where $P_{R,L}=(1 \pm \gamma_5)/2$ and $T^a$ are the Gell-Mann matrices
satisfying ${\mathrm Tr}\, (T^a T^b) = \delta^{ab}/2$.
The $\sigma_{\mu \nu}$ terms are
dimension 5 and absent at tree level in renormalizable theories like the
SM. Hence, they are suppressed by one-loop factors
$\sim \alpha / \pi$. Besides, in the absence of tree level FCN couplings
they are also
suppressed by the GIM mechanism. Thus, these terms are typically small in
renormalizable theories. However, in scenarios with new dynamics near the
electroweak scale effective couplings involving the $t$ quark may be large. On
the other hand, the $\gamma_\mu$ terms can be quite large in principle. Although
rare processes require small FCN couplings
 between light quarks, the top can have
relatively large couplings with the quarks $u$ or $c$, but not with both
simultaneously. In specific models FCN couplings scale with the quark
masses, but this is not general. Simple well-defined models extending the SM
with vector-like fermions can be written fulfilling all precise electroweak data
and saturating the inequalities
\begin{eqnarray}
|X_{tu}^L| \leq 0.28 & , & |X_{tu}^R| \leq 0.14 \,, \nonumber \\
|X_{tc}^L| \leq 0.14 & , & |X_{tc}^R| \leq 0.16
\end{eqnarray}
at 90\% C. L. \cite{papiro13}. It is usually expected that new physics, and in
particular the mass generation mechanism, will show up first in the third family
and thus in the top quark, and large hadron colliders are the best place to
perform a precise measurement of these couplings.
The present 95\% C. L. limits on the top branching ratios at Tevatron are
${\mathrm Br}(t \to Zq) \leq 0.33$,
${\mathrm Br}(t \to \gamma q) \leq 0.032$ \cite{papiro2},
${\mathrm Br}(t \to gq) \leq 0.15$ \cite{papiro3}, which imply
\begin{eqnarray}
X_{tq} & \equiv & \sqrt{|X_{tq}^L|^2+|X_{tq}^R|^2} \leq 0.84 \,, \nonumber \\
\kappa_{tq} & \equiv & \sqrt{|\kappa_{tq}^{(1)}|^2+|\kappa_{tq}^{(2)}|^2} 
\leq 0.778 \,, \nonumber \\
\lambda_{tq} & \equiv & \sqrt{|\lambda_{tq}^{(1)}|^2+|\lambda_{tq}^{(2)}|^2} 
\leq 0.26 \,, \nonumber \\
\zeta_{tq} & \equiv & \sqrt{|\zeta_{tq}^{(1)}|^2+|\zeta_{tq}^{(2)}|^2} 
\leq 0.15 \,.
\label{ec:3}
\end{eqnarray}
Similar limits have been
reported searching for $t\bar q$ production at LEP2 \cite{papiro14}. Relying on
the same decays it has been estimated that LHC with a luminosity of 100
fb$^{-1}$
and future linear colliders will
eventually reduce these bounds to $X_{tq} \leq 0.02$, $\kappa_{tq} \leq 0.015$,
$\lambda_{tq} \leq 0.0035$ \cite{papiro15,papiro16,papiro17,papiro17b}.
In Ref.~\cite{papiro18} the limits on the strong top FCN couplings
have been studied looking at the production of a single top quark
plus a jet at hadron colliders, obtaining $\zeta_{tu} \leq 0.029$,
$\zeta_{tc} \leq 0.11$ in Tevatron Run I.
These will reduce to $\zeta_{tu} \leq 0.0021$,
$\zeta_{tc} \leq 0.0046$ after the first LHC run with a luminosity of
10 fb$^{-1}$.
In the following we investigate in detail what can be
learned from $Zt$ and $\gamma t$ production at Tevatron.

{\bf $\mathbf Zt$ production.}
In general this process manifests as a five fermion final state.
The relatively low
statistics available at Tevatron makes the $\nu \bar \nu j j b$ channel the most
interesting mode due to its branching ratio, $13\%$. The $jjl \nu b$ channel
with a branching ratio of $15\%$ gives less precise results due to its
larger background. We
will only consider $l=e,\mu$ throughout this Letter,
but with an efficient $\tau$
identification the total branching ratio increases by a factor $\sim 3/2$,
improving the significance of
this channel. The $l^+ l^- jjb$ mode has a smaller branching ratio and
the hadronic decay channel $jjjjb$ a larger background, whereas the
three-neutrino channel $\nu \bar \nu l \nu b$ has both a smaller branching ratio
and a larger background. On the other hand, the $b \bar b l \nu b$ and 
$l^+ l^- l \nu b$ modes have smaller branching ratios and backgrounds, and
become the most interesting channels at Tevatron with very high luminosity and
at the LHC. The three charged lepton signal is also characteristic of some gauge
mediated supersymmetry breaking (GMSB) models \cite{papiro9}.
 However, if the origin of such a signal is a top FCN coupling,
 the other decay channels must show up in the ratio
dictated by the $Z$ and $W$ decay rates.

{\em $jjb \etmiss$ signal.}
We discuss first the $\nu \bar \nu jjb$ mode which is the best way of measuring
the $Ztu$ vertex. We will
consider both $\gamma_\mu$ and $\sigma_{\mu \nu}$ terms, but we will assume only
one at a time to be nonzero\footnote{To be definite we fix the ratio $X_{tu}^L /
X_{tu}^R = 4/3$.}.
The samples are generated using the exact matrix element
for the s- and t-channel diagrams $gu \to Zt \to ZWb \to
\nu \bar \nu q \bar q' b$. 
The SM diagrams contributing to $gu \to ZWb$
are much smaller in the phase space region of
interest and suppressed by small mixing angles,
and we neglect them here. We assume all fermions massless except the
top quark. In order to estimate the background we have evaluated four other
processes: ({\em i\/}) $Zjjj$ production using VECBOS \cite{papiro19}
modified to include energy smearing and kinematical cuts; ({\em ii\/}) $Zb
\bar b j$ production, which is much smaller and only important at Run II
when we use $b$ tagging; ({\em iii\/}) the process $gb
\to Wt \to WWb \to l \bar \nu q \bar q' b$, where $l$ is
missed, and ({\em iv\/}) $t \bar t$ production
 with $\bar t (t) \to l \nu b$, with
$l$ and $b$ missed and $t(\bar t) \to q \bar q' b$. We include throughout
this Letter a $K$ factor equal to $1.2$ for all processes \cite{papiro20},
except for $t \bar t$ production where we use $K=1.34$ \cite{papiro21}. 
We use MRST structure functions set A \cite{papiro22} with
$Q^2=\hat s$.
The cross section for both $\gamma_\mu$ or $\sigma_{\mu \nu}$ couplings
is 235 fb at $\sqrt s=1.8$ TeV,
assuming the present upper limits,
$X_{tu} = 0.84$ and $\kappa_{tu}=0.78$.
For $\sqrt s=2$ TeV, the cross sections
increase to 358 and 370 fb, respectively. After generating
signals and backgrounds we imitate the experimental conditions with a Gaussian
smearing of the lepton ($l$), photon ($\gamma$) and jet ($j$) energies,
\begin{eqnarray}
\frac{\Delta E^{l,\gamma}}{E^{l,\gamma}} & = & \frac{20\%}{\sqrt{E^{l,\gamma}}}
\oplus 2\% \,,  \nonumber \\
\frac{\Delta E^{j}}{E^{j}} & = & \frac{80\%}{\sqrt{E^j}} \oplus 5\% \,,
\end{eqnarray}
where the energies are in GeV and
the two terms are added in quadrature. (For simplicity we assume that the
energy smearing for muons is the same as for electrons.) We then apply
detector cuts on transverse momenta $p_T$, pseudorapidities $\eta$ and distances
in $(\eta,\phi)$ space $\Delta R$:
\begin{equation}
p_T^{l,j} \geq 10 {\mathrm ~ GeV} ~,~~ 
p_T^\gamma \geq 16 {\mathrm ~ GeV} ~,~~|\eta^{l,j,\gamma}|  \leq 2 ~,~~
\Delta R_{jj,lj,\gamma l,\gamma j}  \geq 0.4 \,.
\end{equation}
For the $Wt$ and $t \bar t$ backgrounds, we estimate in how many events we
miss the
charged lepton and the $b$ jet demanding that their momenta and pseudorapidities
satisfy $p_T<10$ GeV or $|\eta| > 3$. 

\begin{figure}[ht]
\begin{center}
\mbox{
\epsfig{file=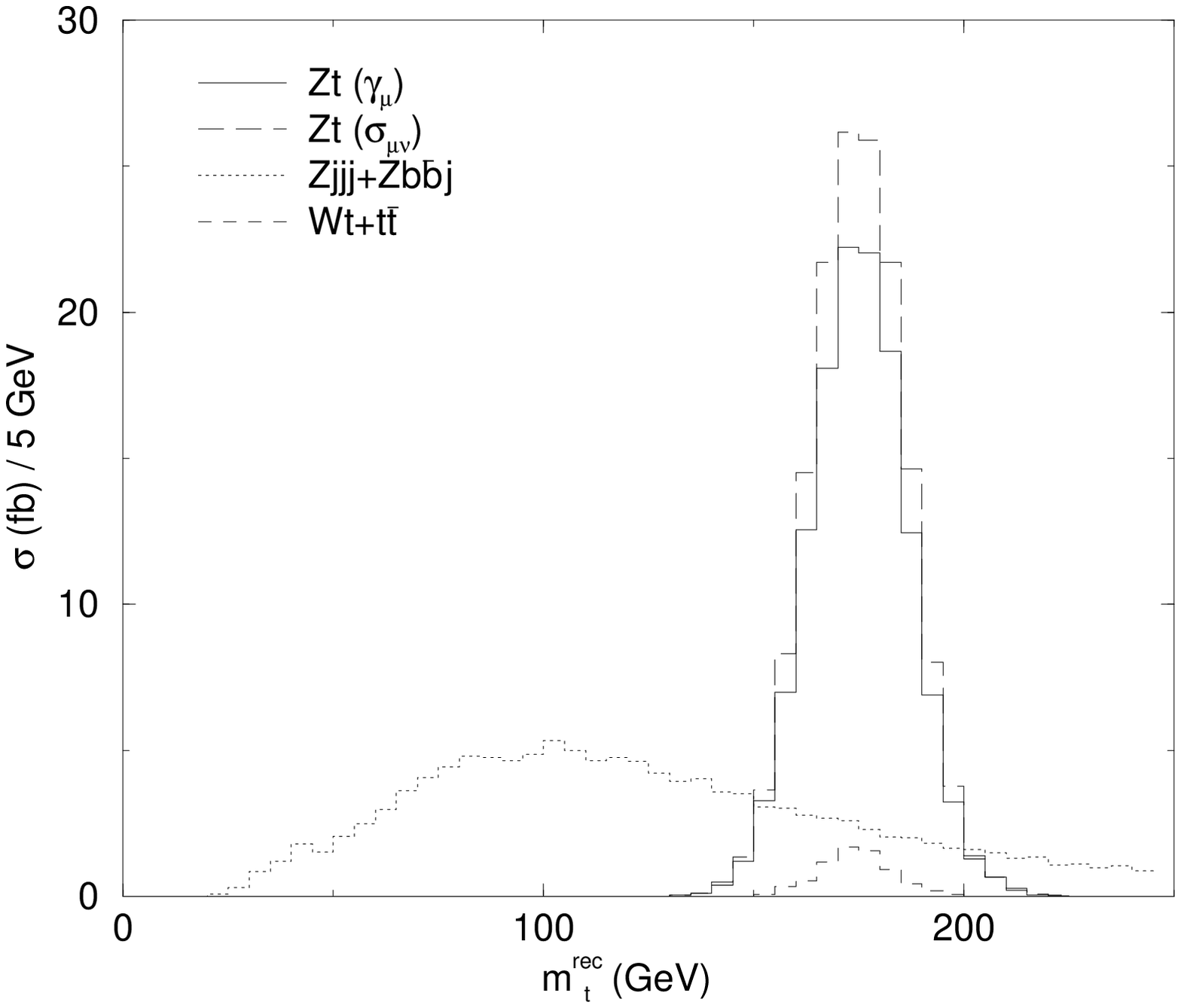,width=10cm}}
\caption{Reconstructed top mass $\mtrec$ distribution before kinematical cuts
for the $\nu \bar \nu jjb$ signal and backgrounds in Tevatron Run II.
\label{fig:mt}}
\end{center}
\end{figure}

\begin{figure}[ht]
\begin{center}
\mbox{
\epsfig{file=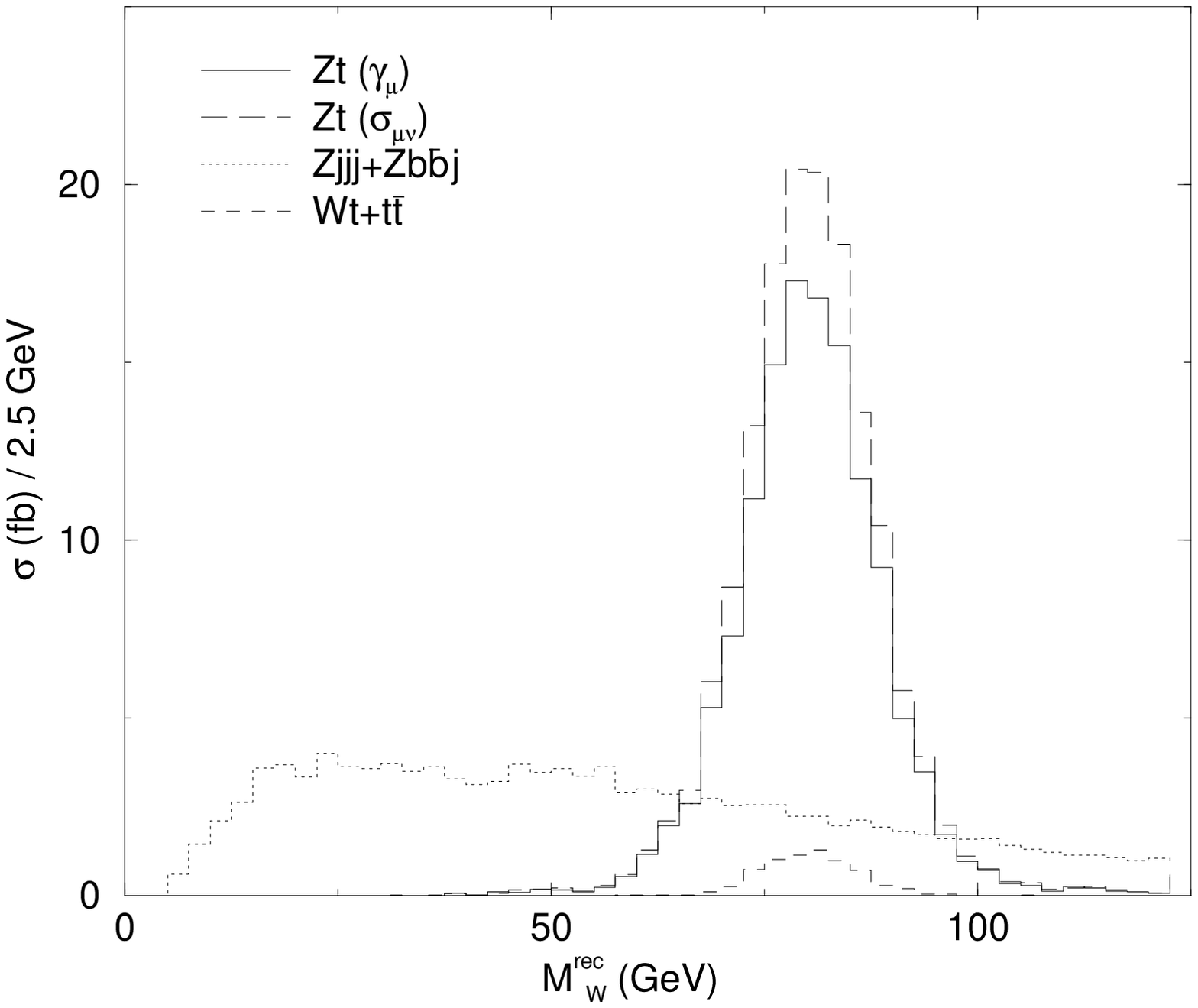,width=10cm}}
\caption{Reconstructed $W$ mass $\mwrec$ distribution before kinematical cuts
for the $\nu \bar \nu jjb$ signal and backgrounds in Tevatron Run II.
\label{fig:mw}}
\end{center}
\end{figure}

\begin{figure}[ht]
\begin{center}
\mbox{
\epsfig{file=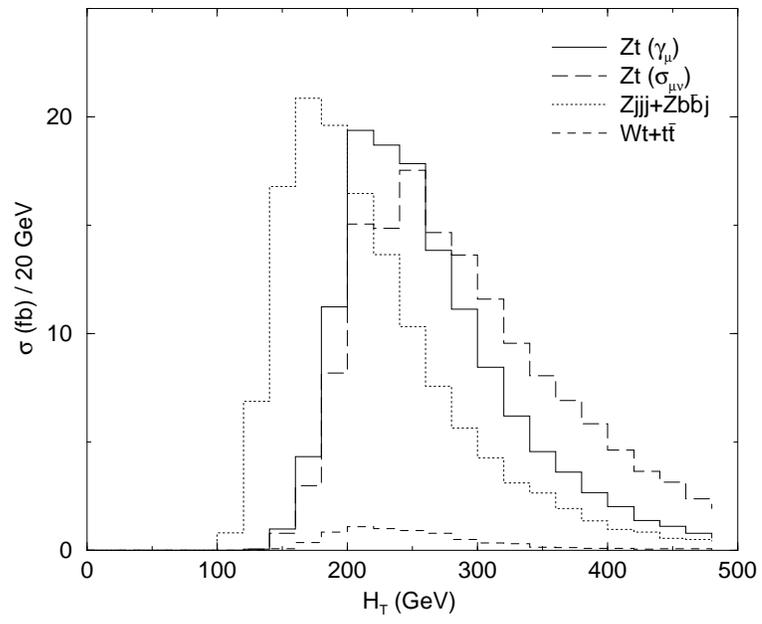,width=10cm}}
\caption{Total transverse energy $H_T$ distribution before kinematical cuts
for the $\nu \bar \nu jjb$ signal and backgrounds in Tevatron Run II.
\label{fig:ht}}
\end{center}
\end{figure}

\begin{figure}[ht]
\begin{center}
\mbox{
\epsfig{file=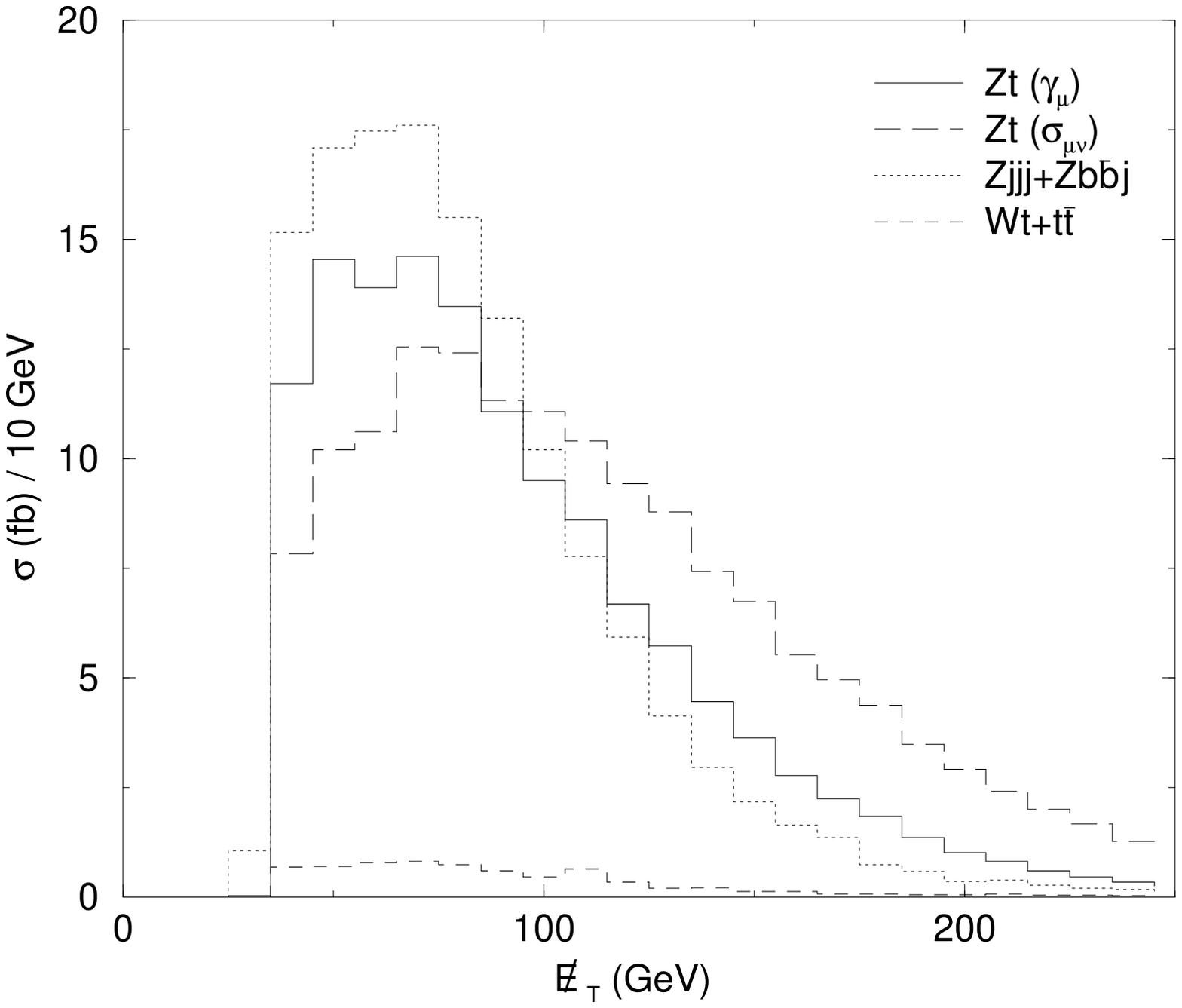,width=10cm}}
\caption{Missing transverse energy $\etmiss$ distribution before kinematical
cuts for the $\nu \bar \nu jjb$ signal and backgrounds in Tevatron Run II.
\label{fig:ptmiss}}
\end{center}
\end{figure}

For the events to be triggered, we require both the signal and background
to fulfil at least one of the following trigger conditions:
\begin{itemize}
\item one jet with $p_T \geq 100$ GeV,
\item one charged lepton with $p_T \geq 20$ GeV and $|\eta| \leq 1$,
\item one photon with $p_T \geq 16$ GeV and $|\eta| \leq 1$,
\item missing energy $\etmiss \geq 35$ GeV and one jet with $p_T \geq 50$ GeV,
\item four jets (including leptons and photons) with $p_T \geq 15$ GeV and 
$\sum p_T \geq 125$ GeV.
\end{itemize}
Finally, for the Tevatron Run II analysis we will take advantage of the
good $b$ tagging efficiency $\sim 60\%$ \cite{papiro25} to require a tagged $b$
jet in the final state. There is also a small probability $\sim 1\%$ that a
jet which does not result from the fragmentation of a $b$ quark
is misidentified as a $b$ jet \cite{papiro25b}.
$b$ tagging is then implemented in the
Monte Carlo routines taking into account all possibilities of $b$
(mis)identification and requiring {\em only} one $b$ jet. This reduces the
signal and the $Wt$ and $t \bar t$ backgrounds by a factor of $\sim 0.6$,
the largest background $Zjjj$ by $\sim 0.03$ and the $Zb \bar b j$ background by
$\sim 0.48$. $b$ tagging is not convenient at Run I because
the number of signal
events and the b-tagging efficiency are small. In Fig.~\ref{fig:mt} we plot the
signal and background distributions for $\mtrec$, the
invariant mass of the
three jets, which is the reconstructed mass of the top quark for the signal.
Obviously in this case $\mtrec$ is not exactly the top mass,
because the $t$ quark is not necessarily on-shell. Besides, we have also
simulated
the detector by smearing the energy. Both effects are in fact comparable.
Obviously, the $\mtrec$ distribution for the $Wt$ and $t \bar t$
backgrounds peaks also around $m_t$. In Fig.~\ref{fig:mw} we plot
the cross section as a function of $\mwrec$, the reconstructed $W$ boson mass.
When we use $b$ tagging, $\mwrec$ is the invariant mass of the other two jets.
If the $b$ is not tagged, $\mwrec$ is defined
as the two-jet invariant mass closest to
the $W$ mass. In this case the third jet is indirectly assigned to a $b$.
$\mwrec$ equals $M_W$ for the signal
and the $Wt$ and $t \bar t$ backgrounds. Another useful variable to discriminate
between signal and background is the total transverse energy $H_T$ in
Fig.~\ref{fig:ht}, which is defined as the scalar sum of the $p_T$'s of all
jets plus $\etmiss$. In Fig.~\ref{fig:ptmiss} we
plot the $\etmiss$ distribution to show
that the possible trigger inefficiency will not change significantly our
results.

To enhance the signal to background ratio we apply different sets of cuts on
$\mtrec$, $\mwrec$, $H_T$ in Runs I and II, and also on
$p_T^b$, the transverse momentum of the $b$ quark,
$p_t^{\mathrm min}$, the minimum transverse momentum of the jets and
$\Delta R_{jj}^{\mathrm min}$, the minimum $\Delta R$ between jets (see
Table~\ref{tab:1}).
The total number of events for Runs I and II with integrated luminosities of
109 pb$^{-1}$ and 2 fb$^{-1}$, respectively, are collected in Table~\ref{tab:2},
using for the signals $X_{tu} = 0.84$ and $\kappa_{tu} = 0.78$.
We observe that the kinematical cuts in Table~\ref{tab:1} are very efficient to
reduce the $Zjjj$ and $Zb \bar b j$ backgrounds, but they do not affect $Wt$
and $t \bar t$. These are in practice irreducible and limit the
usefulness of this decay channel to moderate energies and luminosities. To
derive upper bounds on the coupling constants we use
the prescriptions in Ref. \cite{papiro26}.
These are more adequate than
na\"{\i}ve Poisson statistics when the number of background events is small,
as happens in
our case. (Notice that these prescriptions are similar to those
applied in Ref. \cite{papiro2} to obtain the bounds ${\mathrm Br}(t \to
Zq) \leq 0.33$ and ${\mathrm Br}(t \to \gamma q) \leq 0.032$.)
Unless otherwise stated, all
bounds will be calculated at 95\% C. L. This decay channel gives, if no signal
is observed, $X_{tu} \leq 0.690$, $\kappa_{tu} \leq 0.596$ after Run I
and  $X_{tu} \leq 0.180$, $\kappa_{tu} \leq 0.155$ after Run II.
The expected limit from top decay in Run II is $X_{tu} \leq 0.225$
\cite{papiro15}. Scaling this value with the Run I limits in Eq.~\ref{ec:3} we
estimate $\kappa_{tu} \leq 0.208$ at Run II.

\begin{table}[ht]
\begin{center}
\begin{tabular}{ccc}
Variable & Run I & Run II \\
$\mtrec$ & 155--200 & 155--200 \\
$\mwrec$ & 70--95 & 65--95 \\
$H_T$ & $>180$ & $>160$ \\
$p_T^b$ & & $>20$ \\
$p_T^{\mathrm min}$ & $>20$  \\
$\Delta R_{jj}^{\mathrm min}$ & $>0.6$ 
\end{tabular}
\caption{Kinematical cuts for the $\nu \bar \nu jjb$ decay channel. The masses,
energies and momenta are in GeV. At Run II
we also use $b$ tagging. \label{tab:1}}
\end{center}
\end{table}

\begin{table}[ht]
\begin{center}
\begin{tabular}{ccccc}
 & \multicolumn{2}{c}{Run I} & \multicolumn{2}{c}{Run II} \\
& before & after & before & after \\[-0.2cm]
& cuts & cuts & cuts & cuts \\
$Zt (\gamma_\mu)$ & 16.0 & 10.8 & 261 & 235 \\
$Zt (\sigma_{\mu \nu})$ & 18.1 & 12.4 & 306 & 274 \\
$Zjjj$ & 281 & 7.2 & 199 & 5.2 \\
$Zb \bar b j$ & 4.0 & 0.2 & 74.1 & 2.3 \\
$Wt$ & 0.2 & 0.1 & 3.5 & 3.4 \\
$t \bar t$ & 0.9 & 0.6 & 10.6 & 9.9 \\
\end{tabular} \caption {Number of $\nu \bar \nu jjb$ events before and after
the kinematical cuts in Table~\ref{tab:1} for the $Zt$ signal and backgrounds.
We use $X_{tu}=0.84$ and $\kappa_{tu}=0.78$.  \label{tab:2}}
\end{center}
\end{table}

This process also constrains the strong anomalous top coupling $\zeta_{tu}$.
Again there are
two s- and t-channel diagrams contributing to the signal and a similar
analysis gives
$\zeta_{tu} \leq 0.316$ after Run I and $\zeta_{tu} \leq 0.0824$ after
Run II. These bounds are weaker than the top decay limits $\zeta_{tu} \leq 0.15$
and $\zeta_{tu} \leq 0.04$, respectively \cite{papiro3}, and than the limits
from $jt$ production $\zeta_{tu} \leq 0.029$ and $\zeta_{tu} \leq 0.009$
\cite{papiro18}. One may wonder if it is sensible to use the same cuts for the
$Z$ anomalous terms $\gamma_\mu$ and $\sigma_{\mu \nu}$ and for the strong
$\sigma_{\mu \nu}$ terms. The characteristic $q^\nu$
behaviour differentiates the $\sigma_{\mu \nu}$ from the $\gamma_\mu$ terms and 
manifests differently if the vertex
involves the initial gluon or the final $Z$ boson. However, this
makes little difference for Tevatron energies and we do not distinguish
among the three cases.

{\em $ljjb \etmiss$ signal.}
The $jjl \nu b$ decay channel analysis is carried out in a completely analogous
way. We consider $Wjjj$ and
$Wb \bar b j$ production, which we evaluate with VECBOS, and $Wt$ and $t \bar t$
production, with a $b$ quark missing in the latter,
as backgrounds to our signal (for other single top production processes see
Ref. \cite{papiro28}). To reconstruct the $Z$ boson mass $\mzrec$ 
we use in Run I the two-jet invariant mass closest to the $Z$ mass,
assigning the remaining jet to the $b$. In Run II we require {\em only}
one tagged $b$ defining $\mzrec$ as the invariant mass
of the other two jets\footnote{Requiring only one tagged $b$ reduces
the signal because the $Z$ boson decays to $b \bar b$ $15\%$ of the time.
In this case we have then three $b$'s but we require only one tagged $b$.
The case with more than
one tagged $b$ will be discussed in Ref. \cite{papiro10}.}. We make the
hypothesis that all missing energy comes from a single neutrino with
$p^\nu=(E^\nu,\ptmiss,p_L^\nu)$, and $\ptmiss$ the missing transverse momentum.
Using $(p^l + p^\nu)^2 = M_W^2$ we find two
solutions for $p^\nu$, and we choose that one making the reconstructed top mass
$\mtrec \equiv \sqrt{(p^l +  p^\nu + p^b)^2}$ closest to $m_t$. The complete set
of cuts for Runs I and II is gathered in Table~\ref{tab:3}. In addition
we require $\etmiss > 5$ GeV to ensure that the top mass reconstruction is
meaningful.
The number of events before and after cuts is given in
Table~\ref{tab:4}, taking again for the signals $X_{tu} = 0.84$ and
$\kappa_{tu} = 0.78$.
We observe that although the $Wjjj$ background is one order
of magnitude larger than the $Zjjj$ background of the previous signal,
the cuts are more effective.
This is so in part because $\mtrec$ and $\mzrec$ depend
on different momenta, whereas in the $\nu \bar \nu jjb$ mode
$(\mtrec)^2 = (\mwrec)^2 + 2 p^W \cdot p^b$. In Run I without $b$ tagging
$Wjjj$ is the main background, but at Run II all backgrounds are in practice
comparable. As the only way to get rid of $Wt$ and $t \bar t$ is
distinguishing $Z \to jj$ from $W \to jj$,
we require $\mzrec > 90$ GeV in this Run.
The cuts in Table~\ref{tab:3} are a compromise to reduce
the different backgrounds keeping at the same time
the signal as large as possible.
Thus if no signal is observed,
we obtain from Table~\ref{tab:4}
$X_{tu} \leq 0.838$, $\kappa_{tu} \leq 0.705$ after Run I and 
$X_{tu} \leq 0.275$, $\kappa_{tu} \leq 0.222$ after Run II.

The bounds on $\zeta_{tu}$ are again not competitive with those derived from top
decays and $jt$ production. In this channel we find $\zeta_{tu} \leq 0.374$
(0.119) after Run I (II).

\begin{table}[ht]
\begin{center}
\begin{tabular}{ccc}
Variable & Run I & Run II \\
$\mzrec$ & 80--105 & 90--110 \\
$\mtrec$ & 155--200 & 150--200 \\
$H_T$ & $>240$ & $>240$ \\
$p_T^b$ & $>20$  \\
$\Delta R_{jj}^{\mathrm min}$ & $>0.5$ & $>0.6$
\end{tabular}
\caption{Kinematical cuts for the $jj l \nu b$ decay channel. The masses,
energies and momenta are in GeV. At Run II
we also use $b$ tagging. \label{tab:3}}
\end{center}
\end{table}

\begin{table}[ht]
\begin{center}
\begin{tabular}{ccccc}
 & \multicolumn{2}{c}{Run I} & \multicolumn{2}{c}{Run II} \\
& before & after & before & after \\[-0.2cm]
& cuts & cuts & cuts & cuts \\
$Zt(\gamma_\mu)$ & 17.9 & 9.9 & 259 & 77.2 \\
$Zt(\sigma_{\mu \nu})$ & 19.7 & 12.0 & 284 & 101 \\
$Wjjj$ & 1928 & 13.3 & 1282 & 2.7 \\
$Wb \bar b j$ & 41.6 & 0.2 & 421 & 1.0 \\
$Wt$ & 4.6 & 0.8 & 86.6 & 5.5 \\
$t \bar t$ & 15.2 & 2.7 & 226 & 2.8 
\end{tabular} \caption {Number of $jj l \nu b$ events before and after
the kinematical cuts in Table~\ref{tab:3} for the $Zt$ signal and backgrounds.
We use $X_{tu}=0.84$ and $\kappa_{tu}=0.78$. \label{tab:4}}
\end{center}
\end{table}

{\bf $\mathbf \gamma t$ production.}
This process gives a final state of a photon and three fermions. In this case there
are no $\gamma_\mu$ terms as required by gauge
invariance. Depending
whether the $W$ decays into leptons or hadrons, we have the signal
$\gamma l \nu b$ or $\gamma jjb$.
As in $Zt$ production, we only consider $l=e,\mu$, and
again a good $\tau$ identification will improve our results.
Then the leptonic mode has a branching ratio of
$21\%$, and the hadronic mode $67.9\%$. However, the leptonic mode has a
small background from the SM $\gamma W j$ production,
whereas the hadronic one has a huge
background from $\gamma jjj$ production.

{\em $\gamma lb \etmiss$ signal.}
This signal is again generated using the exact matrix
element for the two s- and t-channel diagrams
$gu \to \gamma t \to \gamma W b \to \gamma l \nu b$ and $\lambda_{tu}=0.26$.
Here we have also neglected the SM diagrams $gu \to \gamma W b$
which are also negligible in the phase space of interest.
The main SM background is $\gamma W j$ production. We consider
$g q_u \to \gamma W q_d$, $g q_d \to \gamma W q_u$ and
$q_u \bar q_d \to \gamma W g$, with $q_u=u,c$ and $q_d=d,s$ (plus
the charge conjugate processes), with the jet misidentified as a
$b$. The true $b$ production from initial $u$ and $c$ quarks is
suppressed by the
Cabibbo-Kobayashi-Maskawa matrix elements $|V_{ub}|^2$ and $|V_{cb}|^2$,
respectively, and is negligible.
To evaluate the background we have first calculated the matrix element for
$g q_u \to \gamma W q_d$, including the eight SM diagrams,
decaying afterwards the $W$ leptonically.
The matrix elements for the other two processes can be
obtained by crossing symmetry. Our result for $\gamma W j$ production at
Tevatron Run I agrees with the cross-section obtained in Ref. \cite{papiro27}.
We use the same detector and trigger cuts as for $Zt$ production.
We also take advantage of $b$ tagging in Run II to reduce the background
(in this process there is only one jet).

\begin{figure}[ht]
\begin{center}
\mbox{
\epsfig{file=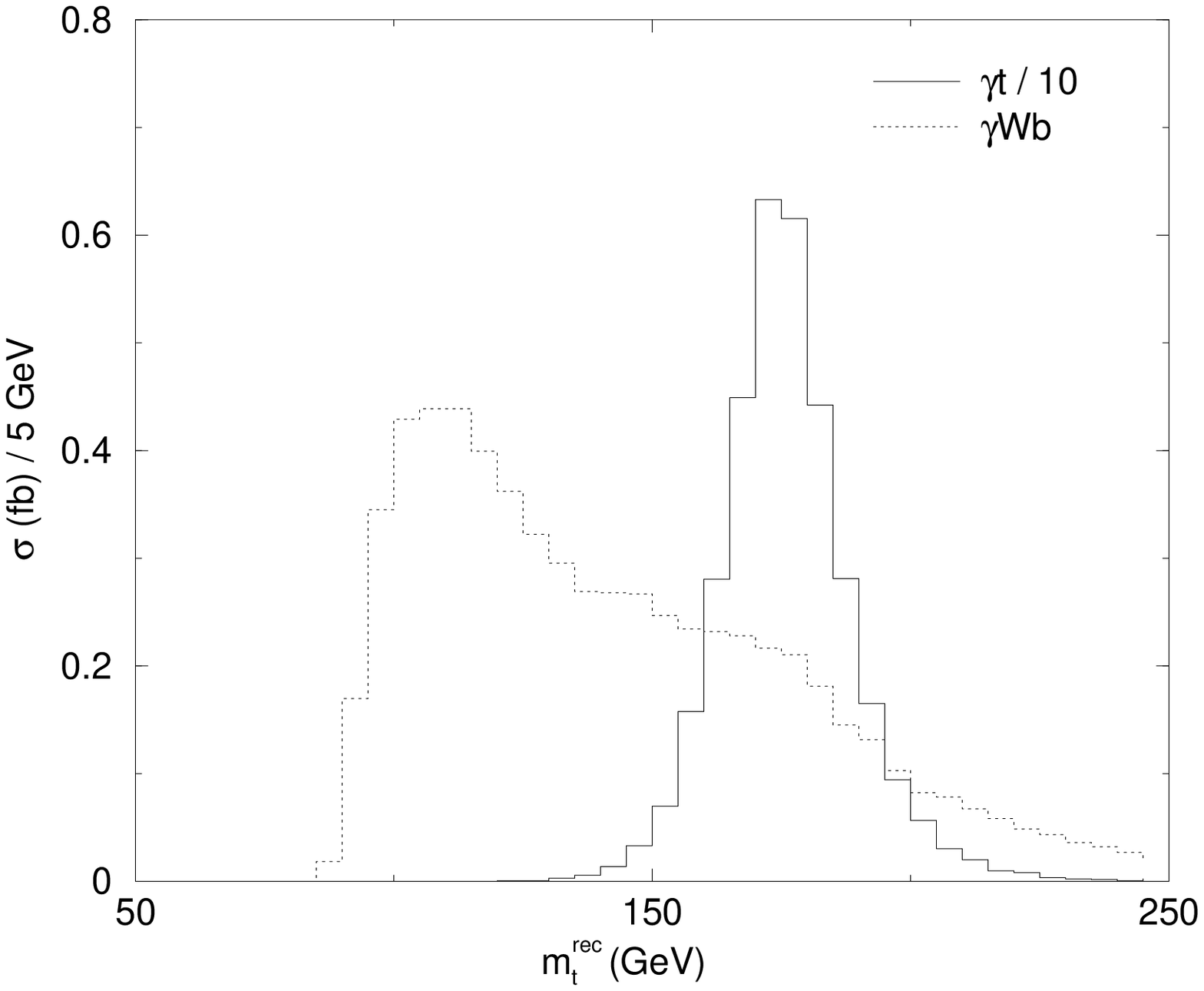,width=10cm}}
\caption{Reconstructed top mass $\mtrec$ distribution before kinematical cuts
for the $\gamma l \nu b$ signal and background in Tevatron Run II. For
comparison the signal distribution has been divided by 10. \label{fig:mt2}}
\end{center}
\end{figure}

\begin{figure}[ht]
\begin{center}
\mbox{
\epsfig{file=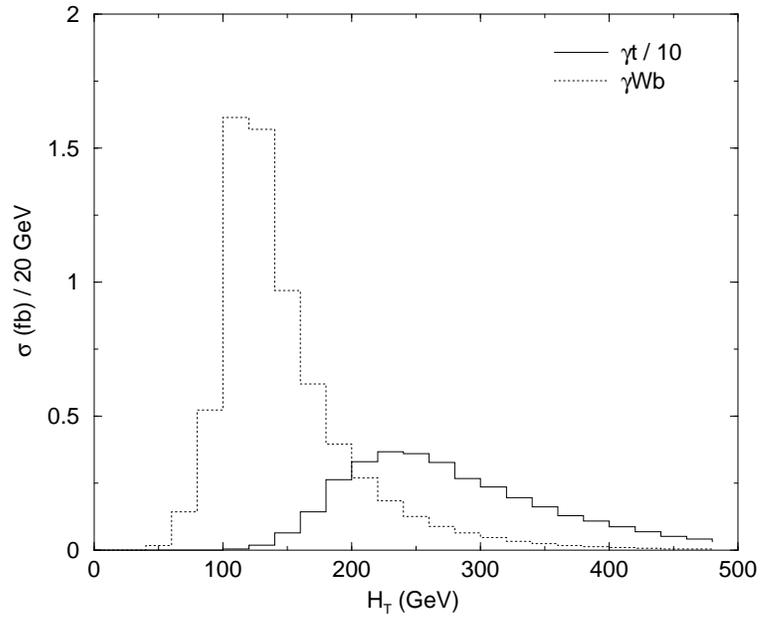,width=10cm}}
\caption{Total transverse energy $H_T$ distribution before kinematical cuts
for the $\gamma l \nu b$ signal and background in Tevatron Run II. For
comparison the signal distribution has been divided by 10. \label{fig:ht2}}
\end{center}
\end{figure}

\begin{figure}[ht]
\begin{center}
\mbox{
\epsfig{file=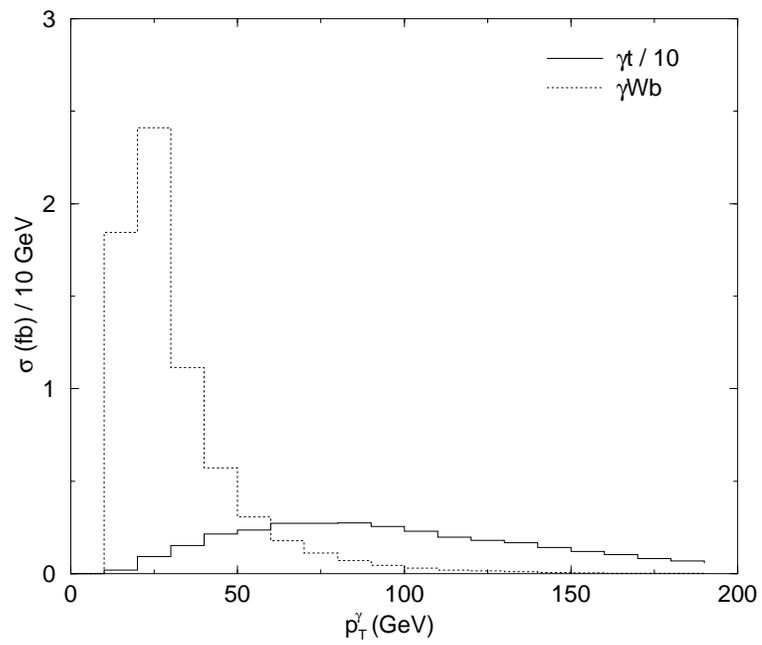,width=10cm}}
\caption{Photon transverse momentum $p_T^\gamma$ distribution before kinematical
cuts for the $\gamma l \nu b$ signal and background in Tevatron Run II. For
comparison the signal distribution has been divided by 10. \label{fig:ptgamma}}
\end{center}
\end{figure}

To improve the signal to background ratio we
perform kinematical cuts on $\mtrec$, which is defined as in the $jj l \nu b$
channel for $Zt$ production (see Fig.~\ref{fig:mt2}).
We also require large $H_T$ (Fig.~\ref{fig:ht2}), $p_T^\gamma$
(Fig.~\ref{fig:ptgamma})
and $E^\gamma$. The
complete set of cuts for Runs I and II is gathered in Table~\ref{tab:5}, where
we have also required $\Delta R_{\gamma W} > 0.4$,
and the number of events in Table~\ref{tab:6}.
Notice that the cuts for Run II are less restrictive than for Run I. This is so
because $b$ tagging alone reduces drastically the background in Run II.
In Run I
the statistics is too low to improve the top decay bound, and we only get
$\lambda_{tu} \leq 0.30$. However with the increase in energy and luminosity
in Run II the characteristic behaviour of the $\sigma_{\mu \nu}$ coupling starts
to manifest and the bound obtained
from this process, $\lambda_{tu} \leq 0.066$, is better than the limit from top
decays, $\lambda_{tu} \leq 0.09$ \cite{papiro16}.

\begin{table}[ht]
\begin{center}
\begin{tabular}{ccc}
Variable & Run I & Run II \\
$\mtrec$ & 150--205 & 140--210 \\
$H_T$ & $>180$ & $>160$ \\
$p_T^\gamma$ & $>40$ & $>30$ \\
$E^\gamma$ & $>50$  \\
\end{tabular}
\caption{Kinematical cuts for the $\gamma l \nu b$ decay channel. The masses,
energies and momenta are in GeV. At Run II
we also use $b$ tagging. \label{tab:5}}
\end{center}
\end{table}

\begin{table}[ht]
\begin{center}
\begin{tabular}{ccccc}
 & \multicolumn{2}{c}{Run I} & \multicolumn{2}{c}{Run II} \\
& before & after & before & after \\[-0.2cm]
& cuts & cuts & cuts & cuts \\
$\gamma t$ & 4.2 & 3.5 & 68.4 & 63.5 \\
$\gamma W q_u$ & 10.3 & 0.5 & 2.5 & 0.3 \\
$\gamma W q_d$ & 10.3 & 0.4 & 2.6 & 0.3 \\
$\gamma W g$ & 39.4 & 1.3 & 8.5 & 0.7 \\
\end{tabular} \caption {Number of $\gamma l \nu b$ events before and after
the kinematical cuts in Table~\ref{tab:5} for the $\gamma t$ signal and
backgrounds. We use $\lambda_{tu}=0.26$. \label{tab:6}}
\end{center}
\end{table}

This process also constrains the strong top FCN vertex. Two s- and t-channel
diagrams contribute to $gu \to \gamma t$ production. Evaluating the exact
matrix element and proceeding as
before we obtain $\zeta_{tu} \leq 0.11$ after Run I and $\zeta_{tu} \leq 0.020$
after Run II. These bounds
are better than those obtained from top decay,
but they are still weaker than the bounds from $tj$ production.

{\em $\gamma jjb$ signal.}
This decay channel has a larger branching ratio than the previous one, but also
a larger background.
In order to evaluate it, we have modified VECBOS to produce photons instead
of $Z$ bosons. This is done introducing a `photon' with a small mass
$m_\gamma=0.1$ GeV and substituting the $Z$ couplings by the photon couplings
everywhere. The total width of such `photon' is calculated to be
$\Gamma_\gamma = 1.73 \cdot 10^{-3}$ GeV, with an $e^+ e^-$ branching ratio
equal to 0.15. We have checked that the results are the same for a heavier
`photon' with $m_\gamma=1$ GeV and $\Gamma_\gamma = 1.73 \cdot 10^{-2}$ GeV.
After detector and trigger cuts, the total number of signal events is
12 at Run I, while the background is huge, 65705 events. However,
with $b$ tagging at Run
II we still can derive a competitive bound on the electromagnetic
anomalous coupling. The
reconstruction of the top and the $W$ mass proceeds as in the $\nu \bar \nu jjb$
signal. The cuts for $\gamma jjb$ are summarized in Table~\ref{tab:7}, 
with $p_T^{\mathrm max}$ the maximum transverse momentum of the three jets. The
number of events before and after cuts can be read
from Table~\ref{tab:8} (for the signal we use $\lambda_{tu}=0.26$).
In this case we derive a bound $\lambda_{tu} \leq 0.088$ similar to that
expected from top decays, but worse that the one obtained in the $ \gamma l
\nu b$ channel.

\begin{table}[ht]
\begin{center}
\begin{tabular}{cc}
Variable & Run II \\
$\mtrec$ & 160--200 \\
$\mwrec$ & 65--95 \\
$H_T$ & $>240$ \\
$p_T^\gamma$ & $>75$ \\
$E^\gamma$ & $>100$  \\
$p_T^{\mathrm max}$ & $>50$ \\
$\Delta R_{jj}^{\mathrm min}$ & $>0.6$
\end{tabular}
\caption{Kinematical cuts for the $\gamma jjb$ decay channel. The masses,
energies and momenta are in GeV. In this Run
we use $b$ tagging. \label{tab:7}}
\end{center}
\end{table}

\begin{table}[ht]
\begin{center}
\begin{tabular}{ccc}
 &  \multicolumn{2}{c}{Run II} \\
& before & after \\[-0.2cm]
& cuts & cuts \\
$\gamma t$ & 192 & 89.9  \\
$\gamma jjj$ & 54290 & 19.1 \\
\end{tabular} \caption {Number of $\gamma jjb$ events before and after
the kinematical cuts in Table~\ref{tab:7} for the $\gamma t$ signal and
background. We use $\lambda_{tu}=0.26$ \label{tab:8}}
\end{center}
\end{table}

This channel also allows to constrain the strong anomalous
coupling $\zeta_{tu}$. Proceeding as before we derive the bound
$\zeta_{tu} \leq 0.048$ after Run II.

In summary, we have shown that $Zt$ and $\gamma t$ production at
large hadron colliders provides a sensitive probe for anomalous FCN top
couplings. At Tevatron energies these processes are sensitive only to $Vtu$
couplings. For $Zt$ production the most interesting channels
are those with $Z \to \nu \bar \nu$ and $W \to q \bar q'$, and
$Z \to q \bar q$ and $W \to l \nu$. For $\gamma t$ production,
both channels $W \to q \bar q'$ and $W \to l\nu$ are significant.
The limits which can be obtained from these signals are quoted in
Table~\ref{tab:9}, together with present and future bounds from
top decays and $jt$
production.

With the increase of the center of mass energy and luminosity at the LHC,
the $l^+ l^- l \nu b$ and $b \bar b l\nu b$ channels will provide
the most precise bounds on $X_{tu}$ and
$\kappa_{tu}$, while the $\gamma l \nu b$ channel will give the strongest bound on
$\lambda_{tu}$ and $\zeta_{tu}$. At
the same time, $Zt$ and $\gamma t$ production from sea $c$ quarks becomes
larger and similar bounds to those from top decays can be obtained. This is
the subject of Ref.~\cite{papiro10}.

\begin{table}
\begin{center}
\begin{tabular}{lcccccccc}
& \multicolumn{4}{c}{Run I} & \multicolumn{4}{c}{Run II} \\
Signal & $X_{tu}$ & $\kappa_{tu}$ & $\lambda_{tu}$ & $\zeta_{tu}$
  & $X_{tu}$ & $\kappa_{tu}$ & $\lambda_{tu}$ & $\zeta_{tu}$ \\
$\nu \bar \nu jjb$ & 0.69 & 0.60 & --- & 0.32
  & 0.18 & 0.15 & --- & 0.082 \\
$jj l \nu b$ & 0.84 & 0.71 & --- & 0.37
  & 0.28 & 0.22 & --- & 0.12 \\
$\gamma l \nu b$ & --- & --- & 0.30 & 0.11
  & --- & --- & 0.066 & 0.020 \\
$\gamma jjb$ & --- & --- & &
  & --- & --- & 0.088 & 0.048 \\
Top decay & 0.84 & 0.78 & 0.26 & 0.15
  & 0.23 & 0.21 & 0.09 & 0.04 \\
$jt$ production& --- & --- & --- & 0.029 & --- & --- & --- & 0.009
\end{tabular}
\caption{Summary of the bounds on the anomalous top couplings in
Eqs.~(\ref{ec:1}),
(\ref{ec:3}) obtained from the main decay channels in single top production in
association with a $Z$ boson or a photon at Tevatron. For comparison we also
quote
the limits from top decay and single top production
plus a jet existing in the literature.
We use dashes to indicate that the process does not constrain the
coupling at tree level. In Run I the $\gamma jjb$ signal gives no significant
bound. \label{tab:9} }
\end{center}
\end{table}

\vspace{1cm}
\noindent
{\Large \bf Acknowledgements}

\vspace{0.4cm} \noindent
We thank W. Giele for helping us with VECBOS and J. Fern\'{a}ndez
de Troc\'{o}niz
and I. Efthymiopoulos for discussions on Tevatron and LHC triggers. We have also
benefited from discussions with F. Cornet, M. Mangano and R. Miquel.
This work was partially supported by CICYT under contract AEN96--1672 and by the
Junta de Andaluc\'{\i}a, FQM101.

\end{document}